\documentclass{article}
\pdfoutput=1

\usepackage{pdfpages}

\title{Mental Disorder Recovery Correlated with Centralities and Interactions on an Online Social Network}

\author{Xinpei Ma \and Hiroki Sayama\\
Center for Collective Dynamics of Complex Systems\\
Department of Systems Science and Industrial Engineering\\
Binghamton University, State University of New York\\
Binghamton, NY 13902-6000, USA\\
xma5@binghamton.edu}

\begin{document}

\maketitle

\begin{abstract}
Recent research has established both a theoretical basis and strong empirical evidence that effective social behavior plays a beneficial role in the maintenance of physical and psychological well-being of people. To test whether social behavior and well-being are also associated in online communities, we studied the correlations between the recovery of patients with mental disorders and their behaviors in online social media. As the source of the data related to the social behavior and progress of mental recovery, we used PatientsLikeMe (PLM), the world\textsc{\char13}s first open-participation research platform for the development of patient-centered health outcome measures. We first constructed an online social network structure based on patient-to-patient ties among 200 patients obtained from PLM. We then characterized patients\textsc{\char13} online social activities by measuring the numbers of “posts and views” and “helpful marks” each patient obtained. The patients\textsc{\char13} recovery data were obtained from their self-reported status information that was also available on PLM. We found that some node properties (in-degree, eigenvector centrality and PageRank) and the two online social activity measures were significantly correlated with patients\textsc{\char13} recovery. Furthermore, we re-collected the patients\textsc{\char13} recovery data two months after the first data collection. We found significant correlations between the patients\textsc{\char13} social behaviors and the second recovery data, which were collected two months apart. Our results indicated that social interactions in online communities such as PLM were significantly associated with the current and future recoveries of patients with mental disorders.\\
Keywords: social media, mental disorder, social network, patient-to-patient network, non-drug treatments, PatientsLikeMe
\end{abstract}

\section*{Introduction}
    Mental health problems (disorders) are medical anomalies that disrupt a person\textsc{\char13}s thinking, feeling, mood, and ability to relate to others, and impair his/her daily functioning \cite{Wakefield:1992dg}. According to the National Alliance on Mental Illness, one in four adults, approximately 57.7 million Americans, experience some sort of mental health disorder every year. Mental disorders might lead to other physical or psychological illnesses and severely interfere with a person\textsc{\char13}s ability to work, study, and entertain \cite{Kessler:2005dg}.\par
    
    Some traditional treatments for mental disorders have been proved insufficient in dealing with the complexities of mental diseases. Somatic and psychotherapeutic treatments are traditional treatment options for mental disorders \cite{Koffmann:2014dg}. Somatic treatments, such as medication and electroconvulsive treatment (ECT), successfully control physical symptoms, but are always associated with side effects like drowsiness, dizziness, muscle spasm and so forth \cite{Leinbaugh:2001dg}. A sudden stop or discontinuation of medicine use is likely to cause relapse. For individuals with moderate or severe mental disorders, both somatic and psychotherapeutic treatments are long-term. The rising cost of mental health services and medicines has put pressure on both patients and mental health providers \cite{Leslie:2014dg}. Nowadays, high relapse rates, side effects, and high costs are three major drawbacks of these treatments. Non-drug treatments, such as interpersonal therapy, peer support groups, and community services, have emerged as cognitive cures for mental illnesses \cite{Fieldhouse:2003dg, Rice:2014dg}. These treatments help patients understand their diseases and manage feelings, thoughts, and actions to improve their mental health conditions \cite{DeRubeis:2001dg,Andersson:2005dg, Perry:2015dg}. Recent reviews of mental health studies pointed out that trust, engagement, communication, and support may strengthen mental functions, and may also buffer negative effects of mental illnesses \cite{Davis:2014dg, Ali:2015dg}.  Since the beneficial role of mutual help and self-help behaviors in the recovery process has received significant attention of both physicians and patients, it is believed that developing effective approaches to investigate these underlying relations has direct implications for the improvement of recovery outcomes \cite{Cohen:2000dg}. \par
    
	Recently, social science research has shown that patients\textsc{\char13} social networking with professionals or other patients could facilitate the development of mutual trust and self-help behaviors \cite{Chow:2008dg}. In some cases, however, face-to-face social interaction may not offer adequate support for patients with mental disorders \cite{Farrell:2003dg}. Limited access to services and increased social stresses for such patients may become obstacles to their social activities, which could easily make them feel isolated \cite{Sadavoy:2004dg}. Over the last decade, however, the proliferation of social media for health promotion has been offering patients opportunities of peer-learning, information sharing and communications \cite{Wells:2007dg,Neiger:2012dg}. Establishing useful and enjoyable social interactions on online social media may soon become a feasible alternative approach for their social life, with low costs, high efficiency and rich diversity. Patients may be able to develop skills to overcome difficulties in communication and recovery, better engage in their disease management processes and brighten their lives through social interactions on those online social media platforms \cite{Yan:2010dg}. \par
    
	The concept of “Health 2.0” has emerged in response to the widespread adoption of web-based platforms for healthcare purposes. These platforms provide patients and healthcare practitioners with electronic channels to store, share and communicate health-related information \cite{VanDeBelt:2010dg}. As those Health 2.0 platforms become an irreplaceable component of today\textsc{\char13}s healthcare systems, more and more patients with mental health problems are inclined to participate in various forms of online communications to gather information and build connections with other patients  \cite{Fox:2009dg}. The Health 2.0 movement also nurtures dozens of startups with creative concepts, which are reforming the healthcare systems globally \cite{Doherty:2008dg,Hawn:2009dg,Myneni:2013dg}. \par
    
Online-based social networking complements face-to-face communication and helps patients improve their self-esteem and social competence \cite{Kummervold:2002dg,Myneni:2013dg, Morris:2015dg}. It encourages patients to be more active in their social environment \cite{Cothrel:1999dg}. For instance, patients may be able to discuss, via online media, their private problems without fear of prejudice or discrimination \cite{Hsiung:2000dg}. Furthermore, online social media may play a complementary role to traditional mental health services and help patients understand their conditions more and take better control over their diseases and behaviors \cite{Frost:2008dg}. For example, while many treatment decisions are still made based on physicians\textsc{\char13} empirical judgments that might not have solid supporting evidence, information sharing via healthcare social media may allow patients to perceive their diseases from other patients\textsc{\char13} point of view, do their own research online, and make their own informed decisions on how to manage their diseases \cite{HERT:2011dg,Frost:2011dg,Wicks:2010dg, Chen:2015dg}. \par
In this paper, we study potential linkages between online social behaviors of patients with mental health disorders and their recovery processes. Patients\textsc{\char13} social network structure and social activities are the two aspects of online social behaviors considered, whose impact on the recovery of mental disorders were investigated. The recovery data were collected twice, at the time of data collection of online social behavior, and then two months later, to examine possible associations between patients\textsc{\char13} online social behaviors and their current and future recovery from mental disorders. \par

\section*{Methods}
\subsection*{Source of Data}

We used PatientsLikeMe (PLM; http://www.patientslikeme.com/), one of the first online communities to encourage patients to share their stories and report their medical histories after receiving therapies \cite{Wicks:2010dg}. PLM has grown to have more than 300,000 members and has gradually expanded to diversified communities involving different kinds of disease such as Parkinson\textsc{\char13}s disease, Multiple Sclerosis (M.S.), HIV, and mood disorder, among other diseases. The authors are not affiliated with PLM and have no financial or other interest in PLM.\par
We sampled, from the Mental Health and Behavior Forum in PLM, 200 patients who (i) have/had mental disorder(s), (ii) have full information about social activities and recovery outcomes from the mental disorder(s), (iii) have registered for more than six months, (iv) have social connections and posted relevant contents or comments to the forum, and (v) do not appear to be a spam, phishing or fake account. We scanned a list of patients who participated in the forum, ranked in the order of their popularity, to collect sample participants. We terminated the sampling when the size of the collected samples that met the above criteria reached 200. The following information was recorded for each patient in our data: (a) online social connections with other existing patients, (b) two online social activities in the forum, and (c) self-assessments of recovery outcomes. \par

\subsection*{Measures}

We calculated six network properties (in-degree, out-degree, betweenness centrality, closeness centrality, eigenvector centrality, and PageRank) by using Gephi \cite{Bastian:2009dg,Borgatti:2009dg}. These network properties are widely used in social network analysis to characterize local and global features of network structures \cite{Bampo:2008dg}. \par

The in- and out- degrees are the numbers of links that go into and out of the node, respectively. In this study, the in-degree represents the number of followers a patient has, which reflects the popularity of the patient in the community. Similarly, the out-degree represents the number of other patients a patient follows, which reflects the willingness and intention of the patient to connect to others \cite{Snijders:2001dg}. Betweenness centrality refers to the probability for a node to be on shortest paths between two other nodes. It is an indicator of the level of control of information flow and influence \cite{Newman:2005dg}. Closeness centrality is the inverse of average distance of all shortest paths from a node to all other nodes in the whole network \cite{Newman:2005dg}. It represents how long and far one node will take to reach other nodes. Both eigenvector centrality and PageRank measure how “important” a node is in the network, taking non-local topological structure into account. The concept underlying those centrality measures assumes that connections from high-importance nodes provide the node more importance than connections from low-importance nodes. PageRank is a variation of eigenvector centrality in which transition probability matrices are used instead of adjacency matrices \cite{Page:1999dg}.\par

For data collection about patients\textsc{\char13} social activities, we recorded (i) how many “helpful marks” patients received and (ii) how many “posts and views” they made per month on the \textit{Mental Health and Behavior Forum} after they registered to PLM. These numbers are the only available numbers online that relate to patients\textsc{\char13} general social activities.\par

The data about patients\textsc{\char13} recovery processes were obtained from patients\textsc{\char13} self-reported health records in PLM. In this website, patients were encouraged to evaluate their physical and psychological feelings and describe their symptoms once a day. We recorded the following five recovery outcomes: mood function, stress, distress, life essentials and symptoms, which are closely related to mental disorders.  In a patient\textsc{\char13}s historical record, these recovery outcomes are visualized as continuous curves or bar graphs (Figures 1, 2 and 3). From these charts, we quantified the rate of change as follows. \par

The curve of mood function shows a patient\textsc{\char13}s ability to regulate his/her mood. Each point on the curve represents how well the patient could control his/her anxiety and mood swings. Similarly, the stress and distress curves show the extent of change in a patient\textsc{\char13}s stressful and distressful feelings, respectively (Figure 1). Dividing a curve into shorter segments and comparing average values of the segments have been commonly used to measure the rate of change for time series data \cite{Ihm:1991dg}. We partitioned each of these curves into two parts of equal length in time, and calculated the average level for each part. The rate of change was then calculated by dividing the difference of two averages (average of the latter part minus average of the former part) by the whole range of the curve (maximum value minus zero) (Equation 1). As a result, the values of those three recovery outcome variables were normalized to the range between -1 and 1 (Figure 4). \par

\begin{equation}
\mathit{Rate\,of\,change} = \frac{\mathit{average\,value\,of\,first\,half}- \mathit{average\,value\,of\,second\,half}}{\mathit{whole\,range\,of\,curve}}
\end{equation}

In PLM, patients\textsc{\char13} life essentials were measured based on patients\textsc{\char13} self-xassessments of life necessities, including sleep, energy, appetite, and sex drive (Figure 2). Similarly to the previous three variables, we separated the records of patients\textsc{\char13} life essentials in two parts of equal length in time. The average level of each part was calculated by converting categorical levels (“Much More”, “More”, “Normal”, “Less”, “Much Less”) to numerical values, and then averaged over all the four life essential variables to create a curve. The computation of the rate of change was done on this curve in the same way as described above, again with the results ranging from -1 to 1.\par

PLM also records general and specific symptoms, which reflect a patient\textsc{\char13}s general mental health conditions as well as particular disease-specific conditions. In this study, we collected some general symptoms (such as fatigue and depressed mood) and some specific symptoms that were closely related to mental disorder conditions (Figure 3). They were converted to numerical values and then aggregated to form a curve, as for the life essentials described above, and then the rate of change was calculated using the same procedure as above. The results were again normalized between -1 and 1.\par
Finally, we roughly estimated the overall recovery outcome of a patient by summing the five recovery variables measured above (Mood Function, Stress, Distress, Life Essentials, and Symptoms). In this study, the five recovery variables are fundamental measurements for patients with mental disorders. Currently, there is no reasonable way to adjust their relative weights, thus we used the simplest possible approach, i.e., a simple sum with equal weights. \par

\subsection*{Analysis}

We first constructed a social network structure among the sampled 200 patients based on their social ties to examine whether or not social networking is associated with their recovery outcomes. The patient-to-patient ties (edges) were established based on “following” relationships in PLM. Since the website suggests a list of other patients with similar conditions or symptoms to each online user, patients tend to follow like-minded patients. In addition, they are also prone to follow popular and helpful ones. In PLM, followers will automatically receive updates from the followed patients, just like in other typical social media. In our study, the direction of a social tie was set from the follower to the followed, representing the direction of attention (i.e., opposite to the direction of information flow). \par

We visualized the patient-to-patient social network structure and calculated important node properties. In the following paragraphs, we conducted a series of statistic analyses to examine the relationship between patients\textsc{\char13} online social behaviors and their recovery outcomes. \par

We calculated Pearson\textsc{\char13}s correlation coefficients between each of the six node properties measured in social network analysis and each of the recovery outcomes (the five recovery outcome variables as well as the overall sum) to find possible associations between network properties and recovery outcomes. The same correlation analysis was also conducted between two online social activities and the recovery outcomes. We did these analyses to identify relevant independent variables that would be incorporated in the following statistical modeling task.\par

Based on the results obtained above, we developed a statistical model of the recovery outcome using multivariable linear regression (we call this part Study I hereafter). The overall recovery outcome was used as the dependent variable and regressed on three explanatory variables of a patient: in-degree, the number of “helpful marks”, and the number of “posts and views”. The reasons of this model setting are threefold. First, the in-degree directly captures the popularity level of a patient and it was found to be significantly correlated with the recovery outcome as well as with other network properties. Second, the numbers of “helpful marks” and “posts and views” are two distinct aspects of patients\textsc{\char13} online social activities, which were also found to be significantly correlated to the recovery outcome. Third, combining measures of social ties and social activities was expected to better represent patients\textsc{\char13} online social behaviors. Up to this point, all the data were collected at a single point in time, with no substantial time delay. We conducted the variation inflation factor (VIF) test to control the issue of multiconlinearity in regression. \par

Furthermore, we re-collected the recovery outcome data for the same 200 patients two months after the initial data collection, and examined associations between online social behaviors and future recovery from mental disorders (we call this part Study II hereafter). The newly collected dependent variable was in the same scale as that of Study I. The same method was applied to measure the rate of change in the recovery outcomes, while the explanatory variables about the patients\textsc{\char13} online social behavior (social ties and activities) were not updated, i.e., they remained at the same values measured two months earlier (i.e., the predicting variables were recorded prior to the measurement of recovery outcomes). Multivariate linear regression was conducted to analyze the relationship between online social behaviors and the overall recovery outcome that were gathered two months apart.\par

The research methods were reviewed and approved by IRB (Protocol Number: 2234-13). The research was conducted with permission of PLM. According to the research protocol reviewed and approved by Binghamton University IRB and to the agreement of data usage with PLM, the researchers are not allowed to share the data with third parties outside the research team. Please contact the corresponding author for more details.\par

\section*{Results}

We first visualized the patients\textsc{\char13} social network structure, which consisted of 200 individuals (nodes) and 981 connections (social ties) (Figure 5). Hu\textsc{\char13}s multilevel graph drawing algorithm was used to lay out the network structure \cite{Hu:2011dg, Walshaw:2001dg}. As a result, a cluster of higher degree nodes gathered in the center of the network, whereas the lower degree nodes were spread across the peripheral area of the graph. \par

The correlation coefficients between six network properties and recovery outcomes are shown in Table 1. The correlations between the in-degree and the three recovery outcomes, including distress reduction, life essentials, and symptoms, were 0.219**, 0.222**, 0.200**, respectively. The correlation between the in-degree and the overall recovery was 0.222**. The results demonstrated that patients who had more incoming social connections (i.e., more followers) experienced greater improvements in feelings of distress, life essentials, and symptoms. PageRank and eigenvector centralities showed similar correlation patterns. As shown in Table 2, the in-degree, eigenvector centrality and PageRank were very strongly intercorrelated. Even though these network measurements capture topologically distinct properties of a network, they are often measuring similar nature of the network and thus show high correlations with each other, especially when the analyzed network has many reciprocated relationships \cite{Valente:2008dg}.  Therefore we chose the in-degree as the social behavior variable to represent all of those three in statistical model building. \par

In the meantime, for out-degree, closeness centrality, and betweenness centrality, their correlations with the recovery outcomes were not statistically significant. The results suggested that following a large number of patients and being close to or in-between other patients in the network were not associated with a significant improvement in recovery outcomes.\par

Table 3 shows the correlation coefficients between the two online social activities and recovery outcomes. The numbers of “helpful marks” patients obtained were strongly correlated with improvements in mood functions, feelings of distress, life essentials, and symptoms as represented in Table 2. Similarly, the number of “posts and views” patients made was also significantly correlated with improvements in the same set of recovery outcomes. Most importantly, both of the two online social activities were significantly correlated with the overall recovery outcome. The numbers of “helpful marks” and “posts and views” were also correlated with each other as well (correlation coefficient is 0.590, \textit{p} $<$ 0.01) (see Table 1 in Supplemental Materials). In the meantime, the “stress” recovery outcome was not significantly correlated with the online social activities. One possible explanation for this observation might be that stress is a reflection of the objective or external conditions felt by patients rather than an implication of subjective control. In short, our results revealed statistically significant correlations between both types of online social activities and all recovery outcomes except stress. \par

Based on the results of correlation analyses described above, we selected the in-degree, the number of “helpful marks”, and the number of “posts and views” as three explanatory variables for the statistical modeling. The overall recovery outcome was used as the sole dependent variable. We conducted two studies for the statistical modeling: Study I used the recovery data collected at the same time as the collection of online social behavior data, while Study II used the recovery data collected two months later. Study I was to model the overall recovery outcome using the three online social behavior variables that were all collected at a single time point. The result is shown in Table 4. The partial \textit{F} statistic was significant (\textit{p} $<$ 0.01), and the overall model explained 15.5\% of variance in the recovery outcomes ($R^{2}$ = 0.155). In this model, only the beta coefficient of the number of “helpful marks” was significant at the 0.05 level. The results showed a statistically significant association between the number of "helpful marks" and recovery outcomes. The variation inflation factor (VIF) test confirmed that there was no multicollinearity among the independent variables (the mean VIF was 1.58, which is far from 10). \par

In Study II, we conducted the same statistical modeling but with the new data of recovery outcomes that were collected for the same 200 patients two months later (while the original social behavior data were still used as is). The result is shown in Table 5. The new model explained 17.3\% of variance in the recovery outcomes ($R^{2}$ = 0.173). The beta coefficients for both “helpful marks” and “posts and views” were significant at the 0.05 level. Overall, the results of Study II revealed that some online social behaviors were significantly correlated with patients\textsc{\char13} recovery outcomes collected two months later. It was suggested that the numbers of “helpful marks” and “views and posts” were associated with the future recovery of patients with mental disorders. \par
In this study, we used SPSS to perform statistical analyses.

\section*{Discussions and Conclusions}

In this study, we investigated possible relationships between the behaviors of patients with mental health disorders in online social media and their recovery outcomes over time. Social network analysis revealed that patients\textsc{\char13} in-degree, eigenvector centrality and PageRank had significant correlations with their recovery outcomes, especially distress reduction, life essentials, and symptoms. The results implied that those high in-degree patients experienced greater reduction of stress, better satisfaction of life essentials and greater alleviation of symptoms, than those who had low in-degrees. Patients\textsc{\char13} online activities, which were characterized by how many “helpful marks” they obtained and how many “posts and views” they made, were also found to be significantly correlated with their recovery outcomes. These findings provide initial evidence that online social behaviors of patients may be positively correlated with their recovery from mental disorders. \par

In order to investigate the relationship between online social activities and recovery outcomes, we constructed two statistical models using the recovery outcome data collected at two time points that were two months apart. Study I used the recovery data collected at the same time as the social behavior data. The result showed that the number of “helpful marks” was correlated with the patients\textsc{\char13} recovery outcomes. Study II used another set of recovery data collected two months later. The result showed that the numbers of both “helpful marks” and “views and posts” were significantly correlated with the patients\textsc{\char13} recovery outcomes. Study I and Study II suggested that patients\textsc{\char13} social interactions in online social media were strongly correlated with their current and future recovery from mental disorders.\par

We note that there are several important limitations in this study. First, we collected the data through scanning the \textit{Mental Health and Behavior Forum}, starting with the top ranked patients, and terminated the sampling when the sample size reached the capacity of our data collection/processing capability. This must have resulted in underrepresentation of patients who did not engage much in social activities. Second, although we attempted to control some of demographic variables (e.g., we confirmed that the ages of sampled patients were nearly evenly distributed, with the mean age being 40.6 years old), we did not have enough information to control other variables such as gender, disease type, residential locations, social/ economic status, education level, etc. Third, only the connections within these 200 patients were considered to construct social network structure, and link weights were ignored in this process. To fully capture and represent the social environment for each patient, other links to/from the outside of this sample group as well as the variation of link weights should be included in the social network analysis. In order to overcome these limitations, a more systematic, fully data-driven research should be conducted. Finally, both distributions of patients\textsc{\char13} recovery outcomes in Study I and Study II approximated normal distributions with mean 0.436 and 0.454, respectively (see Figure 1 and Figure 2 in Supplemental Materials). The results indicated that, among the 200 participants, the majority of patients reported that they had improvements in their recovery outcomes after using PLM. In this study, we did not seek to quantify the potential bias of online self-reported data, which might limit the interpretability of the obtained results. Addressing this problem and improving the accuracy of the analysis requires further effort.\par

To sum up, in this paper, we investigated the association between typical online social behaviors and recovery outcomes of mental disorders.  Even though this study has produced some evidence of possible links between online social interactions and mental health improvement, the important issue regarding precise mechanisms and causal pathways, through which social activities affect mental health outcomes and/or vice versa, still remains unclear. In order to obtain a conclusive answer to the question about how online social behavior and mental health improvement are causally linked, a much larger-scale longitudinal study (ideally with randomized control experiments) would be necessary. In the meantime, we believe that our finding that online social behaviors are strongly linked to patients\textsc{\char13} current and future recovery still has merit by itself, even without full understanding of its causality. \par

\bibliographystyle{plain}
\bibliography{reference}



\section*{Supplementary material}

**Figure 1:** *Patient's Recovery Records about Mood Function, Distress and Stress*

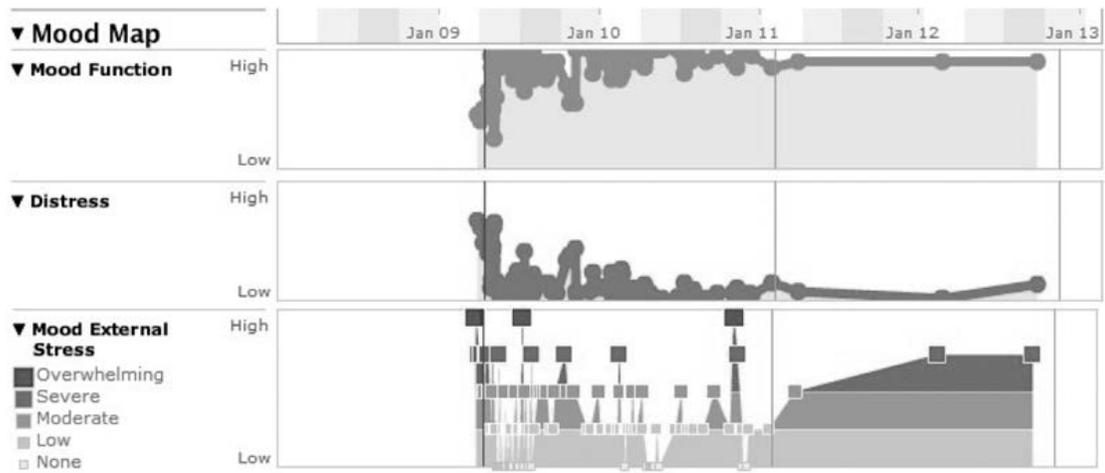

**Figure 2:** *Life Essentials*

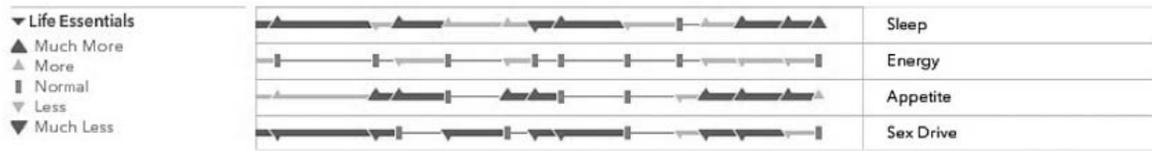

**Figure 3:** *Symptoms*

| | Jan 09 | Jan 10 | Jan 11 | Jan 12 | Jan 13 | |
|---|---|---|---|---|---|---|
| | | | | | | ▼ **General Symptoms** |
| | | | ◆——○ | | | ► Anxious mood |
| | | | ◆——◆ | | | Depressed mood |
| | | | ◆——◆ | | | Fatigue |
| | | | ○——◆ | | | Insomnia |
| | | | ◆——◆ | | | Pain |
| | | | | | | ▼ **Bipolar I Disorder** |
| | | | ◆——◆ | | | Delusions |
| ━━━━━━━━━━━━━━━━●━◆——○ | | | | | | Excitability |
| ─────────○─────○ | | | | | | Flight of ideas |
| | | | ◆——◆ | | | Grandiose thinking |
| | | | ◆——◆ | | | Hallucinations |
| | | | ○——○ | | | Irritability |
| | | | ◆——◆ | | | Paranoia |
| | | | | | | ▼ **ADHD (Attention Deficit/Hypera** |
| | | | | ◆ | | Hyperactivity |
| | | | ◆——◆ | | | Impulsive behaviors |
| | | | ○——○ | | | Inattention |
| | | | | | | ▼ **Migraine** |
| | | | | ◆ | | Light sensitivity (photophobia) |
| | | | | ○ | | ► Migraine headaches |
| | «‹‹‹◆‹◆‹◆‹‹‹‹‹‹‹◆‹◆‹‹‹◆◆◆◆◆◆◆— | | | ◆ | | Nausea |

▼ **Symptoms**

Severity of symptoms
- ■ None
- ■ Mild
- ■ Moderate
- ■ Severe

► Clicking the arrow will display treatments for that symptom
- ■ Prescription Drug
- ■ Equipment

**Figure 4:** *Schematic Diagram of How to Calculate the Rate of Change of A Recovery Curve. This diagram shows a time series of a patient's mood function changes.*

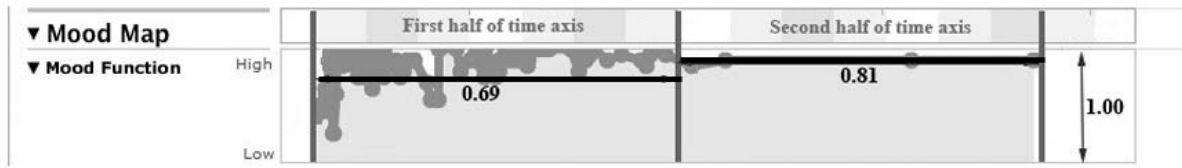

**Figure 5:** *Social Network Structure of 200 Sampled Patients with Mental Disorders. Nodes are shaded according to their degrees.*

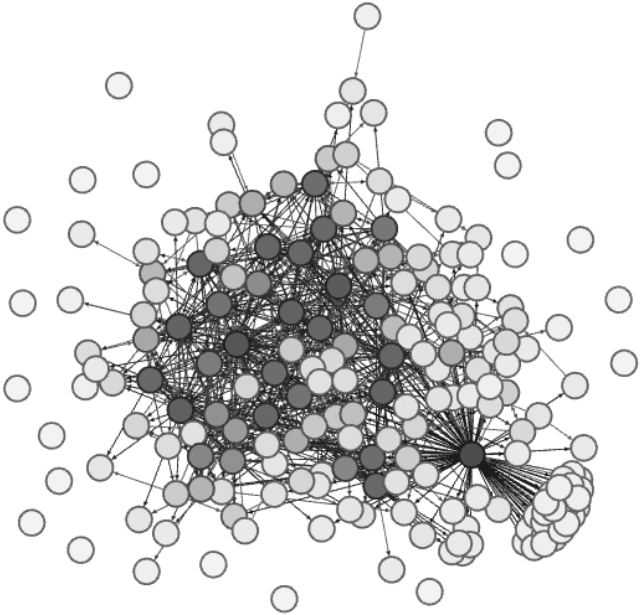

| Correlation Matrix | In-Degree | Out-Degree | Closeness Centrality | Betweenness centrality | Eigenvector centrality | PageRank |
|---|---|---|---|---|---|---|
| Mood Function | 0.08 | 0.067 | -0.071 | 0.067 | 0.094 | 0.902 |
| Stress | -0.014 | 0.036 | -0.087 | -0.085 | -0.024 | -0.024 |
| Distress | 0.219** | 0.119 | -0.015 | 0.181* | 0.223** | 0.224** |
| Life Essentials | 0.222** | 0.119 | 0.042 | 0.109 | 0.235** | 0.196** |
| Symptoms | 0.200** | 0.134 | -0.069 | 0.143 | 0.216* | 0.208** |
| Overall Recovery | 0.222** | 0.125 | -0.035 | 0.144 | 0.234** | 0.208** |

**Table 1:** *Correlation Coefficient Matrix between Six Node Properties and Five Recovery Outcomes (Overall recovery is represented by the sum of all variables listed above)*
** Correlation is significant at the 0.01 level
*Correlation is significant at the 0.05 level

| | In-Degree | Out-Degree | Closeness Centrality | Betweenness Centrality | Eigenvector Centrality | PageRank |
|---|---|---|---|---|---|---|
| **In-Degree** | 1 | | | | | |
| **Out-Degree** | .442** | 1 | | | | |
| **Closeness Centrality** | .286** | .279** | 1 | | | |
| **Betweenness Centrality** | .557** | .732** | .177** | 1 | | |
| **Eigenvector Centrality** | .975** | .424** | .322** | .509** | 1 | |
| **PageRank** | .941** | .494** | .265** | .648** | .917** | 1 |

**Table 2:** *Correlation Coefficient Matrix of Six Network Properties*
** Correlation is significant at the 0.01 level
* Correlation is significant at the 0.05 level

| Correlation Matrix | Helpful Marks | Posts & Views |
|---|---|---|
| Mood Function | 0.147* | 0.141* |
| Stress | -0.013 | 0.026 |
| Distress | 0.216** | 0.147** |
| Life Essentials | 0.316** | 0.226** |
| Symptoms | 0.209** | 0.177* |
| Overall Recovery | 0.279** | 0.242** |

**Table 3:** *Correlation Coefficients between Online Social Activities and Recovery Outcomes (Overall recovery is represented by the sum of all variables listed above)*
\*\* Correlation is significant at the 0.01 level
\*Correlation is significant at the 0.05 level

| | Beta Coefficient | Standard Error | t-Value | p-value |
|---|---|---|---|---|
| Intercept | -0.066 | 0.107 | -0.615 | 0.539 |
| In-degree | 0.070 | 0.040 | 1.759 | 0.080 |
| Helpful Marks | 0.152 | 0.077 | 2.000 | 0.047* |
| Posts and Views | 0.130 | 0.069 | 1.881 | 0.062 |

**Table 4:** *Summery Statistics of Multivariate Linear Regression in Study I*
**R2:** 0.155
**F-statistic:** 11.037 ($p < 0.01$)
**Residual standard error:** 0.516

| | Beta Coefficient | Standard Error | t-Value | p-value |
|---|---|---|---|---|
| Intercept | -0.093 | 0.111 | -0.839 | 0.403 |
| In-degree | 0.066 | 0.040 | 1.619 | 0.107 |
| Helpful Marks | 0.161 | 0.079 | 2.055 | 0.041* |
| Posts and Views | 0.165 | 0.071 | 2.312 | 0.022* |

**Table 5:** *Summery Statistics of Multivariate Linear Regression In Study II*
$R^2$: 0.173
**F-statistic:** 12.574 ($p < 0.01$)
**Residual standard error:** 0.531



\end{document}